\documentclass[11pt]{article}

\usepackage{cite}

\usepackage[dvips]{graphicx}
\usepackage{bm}
\usepackage{amsmath,amssymb}
\usepackage{comment}

\newcommand{\nt}{\nonumber\\}
\newcommand{\nn}{\nonumber}

\newcommand{\eps}{\epsilon}

\newcommand{\alp}{\alpha}

\newcommand{\lam}{\lambda}
\newcommand{\sig}{\sigma}

\newcommand{\lag}{\langle}
\newcommand{\rag}{\rangle}

\newcommand{\back}{\!\!\!\!\!\!}

\newcommand{\cA}{{\cal A}}

\newcommand{\cN}{{\cal N}}
\newcommand{\cL}{{\cal L}}
\newcommand{\cM}{{\cal M}}

\newcommand{\umu}{{\udl \mu}}
\newcommand{\unu}{{\udl \nu}}

\newcommand{\ve}{{\vec e}}
\newcommand{\vm}{{\vec m}}
\newcommand{\vn}{{\vec n}}
\newcommand{\vl}{{\vec l}}

\newcommand{\vy}{{\vec y}}

\newcommand{\vlam}{{\vec \lambda}}
\newcommand{\vpi}{{\vec \pi}}

\newcommand{\udl}{\underline}

\newcommand{\ba}{\begin{eqnarray}}
\newcommand{\ea}{\end{eqnarray}}
\newcommand{\ban}{\begin{eqnarray*}}
\newcommand{\ean}{\end{eqnarray*}}
\newcommand{\sss}[1]{\subsubsection*{#1}}

\begin{document}
\begin{titlepage}
\begin{flushright}
KEK-TH-1439
\end{flushright}

\vskip 16mm

\begin{center}
{\bf\LARGE
D$p$-branes, NS5-branes and U-duality from\\[+7pt]
nonabelian (2,0) theory with Lie 3-algebra}

\vskip 16mm

{\Large
 Yoshinori Honma\footnote
{E-mail address: yhonma@post.kek.jp},
 Morirou Ogawa\footnote
{E-mail address: morirou@post.kek.jp}
 and Shotaro Shiba\footnote
{E-mail address: sshiba@post.kek.jp}
}

\vskip 7mm

{\it\large
Institute of Particle and Nuclear Studies,\\
High Energy Accelerator Research Organization (KEK)\\
and\\
Department of Particle and Nuclear Physics,\\
The Graduate University for Advanced Studies (SOKENDAI)\\[+7pt]
Oho 1-1, Tsukuba, Ibaraki 305-0801, Japan\\
\noindent{ \smallskip }\\
}
\end{center}

\vskip 16mm

\begin{abstract}
We derive the super Yang-Mills action of D$p$-branes on a torus $T^{p-4}$
from the nonabelian $(2,0)$ theory with Lie 3-algebra~\cite{Lambert:2010wm}.
Our realization is based on Lie 3-algebra with pairs of Lorentzian metric generators.
The resultant theory then has negative norm modes,
but it results in a unitary theory by setting VEV's of these modes.
This procedure corresponds to the torus compactification,
therefore by taking a transformation which is equivalent to T-duality,
the D$p$-brane action is obtained.
We also study type IIA/IIB NS5-brane and Kaluza-Klein monopole systems 
by taking other VEV assignments.
Such various compactifications can be realized in the nonabelian $(2,0)$ theory, 
since both longitudinal and transverse directions can be compactified,
which is different from the BLG theory.
We finally discuss U-duality among these branes, and show that
most of the moduli parameters in U-duality group are recovered.
Especially in D5-brane case, the whole U-duality relation is properly reproduced.
\end{abstract}

\end{titlepage}

\setcounter{footnote}{0}

\section{Introduction}





Triggered by the pioneer papers~\cite{Bagger:2006sk,Bagger:2007jr,Gustavsson:2007vu},
fruitful developments about the multiple M2-branes has been achieved.
The novelty is the appearance of Lie 3-algebra $[T^a,T^b,T^c]=f^{abc}{}_d T^d$
as gauge symmetry,
and the theory based on this algebra has appropriate
symmetries as the effective theory of multiple M2-branes.
This is called Bagger-Lambert-Gustavsson (BLG) theory.
For the concrete expressions of Lie 3-algebra,
it is known that the following theories with maximal supersymmetry
can be derived from the original BLG theory:
$\cA_4$ BLG theory for two M2-branes~\cite{Bagger:2007jr},
Lorentzian BLG theory for multiple D2-branes~\cite{Mukhi:2008ux,Ho:2008ei},
extended Lorentzian BLG theory for multiple D$p$-branes ($p>2)$~\cite{Ho:2009nk,Kobo:2009gz},
and Nambu-Poisson worldvolume theory for a single M5-brane~\cite{Ho:2008nn,Ho:2008ve} or finite number of multiple M2-branes~\cite{Chu:2008qv}.
Another approach to construct the action of multiple M2-branes is given by~\cite{Aharony:2008ug}, and this Aharony-Bergman-Jafferis-Maldacena
 (ABJM) theory describes an arbitrary number $N$ of multiple M2-branes on an orbifold ${\mathbb{C}}^4/{\mathbb{Z}}_k$.
This theory has $U(N)\times U(N)$ gauge symmetry and only in special cases
it can have maximal supersymmetry.
In fact, ABJM theory in a certain scaling limit reproduces Lorentzian BLG theory,
and the latter theory can be reduced to the 3-dim super Yang-Mills theory
through the new kind of Higgs mechanism~\cite{Mukhi:2008ux}.
Therefore, the relation between M2-branes and D2-branes are clarified in the viewpoint of the worldvolume theories~\cite{Honma:2008un,Honma:2008jd,Honma:2008ef} (see also
\cite{Antonyan:2008jf,Chu:2010fk}).
In addition, when we start from the extended Lorentzian BLG theory~\cite{Ho:2009nk,Kobo:2009gz} or the orbifolded ABJM theory~\cite{Hashimoto:2008ij,Honma:2009bx},
we obtain D$p$-branes whose worldvolume is a flat torus $T^{p-2}$ bundle over the membrane worldvolume.
In these cases, the moduli of torus compactification of M-theory is properly realized,
and the U-duality transformation can be expressed in terms of Lie 3-algebra or the quiver of Lie groups.

On the other hand, there has been a long time mystery about M5-brane. 
It is known that the low energy dynamics of M5-brane is described by 6-dim $(2,0)$ SCFT,
and that the field contents are five scalars, a spinor and a self-dual 2-form field.
However, the covariant description of the self-dual field is not easy,
and so only the covariant action of single M5-brane is known~\cite{Pasti:1997gx,Bandos:1997ui,Howe:1997fb}.
For the multiple M5-brane dynamics, it has not been known even in the level of the equations of motion.
Recently, however, Lambert and Papageorgakis~\cite{Lambert:2010wm} proposed a set of equations of motion of the nonabelian $(2,0)$ theory by using the Lie 3-algebra,
which may shed light on the underlying cause of the mystery.
Starting from the supersymmetry transformation of the multiple D4-branes theory,
they conjectured that of the nonabelian $(2,0)$ theory.
Through the construction, they introduce an auxiliary field which doesn't appear in the abelian case.
Although this theory seems simply reduced to 5-dim super Yang-Mills theory and might be nothing more than the reformulation of D4-brane theory, 
this must be the first step toward the covariant description of multiple M5-branes and further investigations need to be done.

In this paper, we compactify the transverse/longitudinal directions of the nonabelian $(2,0)$ theory to obtain the 
various brane theories,
which we suppose shows that this theory has something more than D4-brane theory as a model of multiple M5-branes.
First we rewrite the equations of motion of nonabelian $(2,0)$ theory by using the central extension of
Lorentzian Lie 3-algebra.
Although the centers of the Lie 3-algebra cause ghost fields in the theory,
they can be removed by the Higgs mechanism.
In this mechanism, we assign VEV's to these ghosts which determine 
the directions and the radii of the torus compactification.
If we choose suitable VEV's,
we obtain 5-dim maximally super Yang-Mills theory with Kaluza-Klein tower.
In this case, these Kaluza-Klein modes can be associated only with the subspace of 10-dim spacetime where superstring theory lives and not with the M-theory direction,
which is consistent with~\cite{Lambert:2010iw}.
Then, by a rearrangement of fields corresponding to T-duality~\cite{Taylor:1996ik},
these Kaluza-Klein modes can be interpreted as the winding modes of higher dimensional branes,
and we finally obtain D$p$-branes $(p>4)$.
For other choices of VEV's, we can also obtain type IIA/IIB NS5-brane 
and the Kaluza-Klein monopole systems.
We also discuss how the U-duality transformation~\cite{Hull:1994ys}
is realized among these branes.
Although the situation is very similar to the torus compactification of
multiple M2-brane case~\cite{Kobo:2009gz}, it is highly nontrivial that 
at this time we can also realize the S-duality relation between D5-branes and NS5-branes.
And in the case of D5- and D6-branes, we can recover the whole moduli parameters of U-duality.
However, we find that the moduli space is not fully realized in the higher dimensional D$p$-brane cases,
and it is just conceivable that we need to modify the formalism.

This paper is organized as follows.
In \S\,\ref{sec:M5}, we briefly review the formulation of the nonabelian $(2,0)$ theory
with Lie 3-algebra.
In \S\,\ref{sec:Dp}, we explicitly show how D$p$-brane system emerges from the nonabelian $(2,0)$ theory.
Furthermore, by changing the direction of M-theory compactification and that of taking T-duality from the D$p$-brane case, 
we obtain type IIA/IIB NS5-brane system in \S\,\ref{sec:NS5}.
In \S\,\ref{sec:gen}, we make complementary discussion needed to recover the moduli parameters of U-duality,
and also comment on type IIA/IIB Kaluza-Klein monopole system.
In \S\,\ref{sec:U-duality}, we investigate the realization of U-duality transformation in the resultant D$p$\,/NS5-brane actions.
Finally, we conclude in \S\,\ref{sec:Concl}.

\paragraph{Note added}
In the final stage of writing this paper, we were informed of 
the related work by Shoichi Kawamoto, Tomohisa Takimi and Dan Tomino~\cite{Kawamoto:2011}.
We would like to thank them for useful discussions and coordinating the date of submission each other.

\section{Nonabelian (2,0) theory with Lie 3-algebra}
\label{sec:M5}


The nonabelian $(2,0)$ theory proposed in~\cite{Lambert:2010wm}
is given by the following set of equations of motion (EOM)
\ba\label{EOM}
D_\mu^2X_a^I
-\frac{i}{2}[C^\mu,\bar\Psi,\Gamma_\mu\Gamma^I\Psi]_a
-[C^\mu,X^J,[C_\mu,X^J,X^I]]_a
\!&=&\!0\nt
\Gamma^\mu D_\mu\Psi_a+\Gamma_\mu\Gamma^I[C^\mu,X^I,\Psi]_a
\!&=&\!0\nt
D_{[\mu}H_{\nu\rho\sig]a}
+\frac14\eps_{\mu\nu\rho\sig\lam\tau}[C^\lam,X^I,D^\tau X^I]_a
+\frac{i}{8}\eps_{\mu\nu\rho\sig\lam\tau}[C^\lam,\bar\Psi,\Gamma^\tau\Psi]_a
\!&=&\!0\nt
\tilde F_{\mu\nu}{}^b{}_a-C^{\rho}_c H_{\mu\nu\rho,d}f^{cdb}{}_a
\!&=&\!0\nt
D_\mu C^\nu_a
\!&=&\!0\,,
\ea
and constraints
\ba
\label{cons}
C_c^\mu D_\mu X_d^I f^{cdb}{}_a
=C_c^\mu D_\mu \Psi_d f^{cdb}{}_a
=C_c^\mu D_\mu H_{\nu\rho\sig,d}f^{cdb}{}_a
=C_c^\mu C_d^\nu f^{cdb}{}_a
\!&=&\!0\,.
\ea
This theory has 6-dim $\cN=(2,0)$ supersymmetry and nontrivial gauge symmetry,
so this formulation is expected to be a new approach to understand the 
multiple M5-brane dynamics.
Here
the indices $I=6,\cdots,10$ specify the transverse directions of M5-branes 
and $\mu,\nu=0,\cdots,5$ indicate the longitudinal directions.
$a,b,\cdots$ denote the gauge indices.

The field contents are as follows:
$X^I_a$ are scalar fields, $\Psi_a$ is a spinor field,
$A_{\mu,ab}$ is a gauge field,
and $C^\mu_a$ is a new auxiliary field. 
It is well known that
the 6-dim $\cN=(2,0)$ tensor multiplet contains the 2-form field $B_{\mu\nu,a}$
besides $X^I_a$ and $\Psi_a$.
In this theory, only its field strength
$H_{\mu\nu\rho,a}=3\partial_{[\mu} B_{\nu\rho]a}$ appears and it satisfies the self-dual condition
\ba\label{dual}
H_{\mu\nu\rho,a}=\frac{1}{3!}\eps_{\mu\nu\rho\sig\lam\tau}H^{\sig\lam\tau}{}_a\,.
\ea
The covariant derivative of the fields $\Phi=X^I,\,\Psi,\,H_{\mu\nu\rho},\,C^\mu$ is defined by
\ba
(D_\mu\Phi)_a:=\partial_\mu\Phi_a-if^{cdb}{}_a A_{\mu,cd}\Phi_b\,,
\ea
where the notation is slightly different from the original one~\cite{Lambert:2010wm},
so that the gauge field $A_{\mu,ab}$ becomes Hermitian.

\paragraph{Lie 3-algebra}

In general, Lie 3-algebra is defined with the totally antisymmetric 3-bracket and the inner product
\ba
{}[T^a,T^b,T^c]=f^{abc}{}_d T^d\,,\quad
\lag T^a,T^b\rag=h^{ab}\,,
\ea
where $f^{abc}{}_d$ is a structure constant and $h^{ab}$ is a metric.
For the closure of gauge transformation, the structure constant must satisfy the fundamental identity
\ba\label{fund}
 f^{abc}{}_f f^{def}{}_g
+f^{abd}{}_f f^{ecf}{}_g
+f^{abe}{}_f f^{cdf}{}_g
=f^{cde}{}_f f^{abf}{}_g\,.
\ea
Also, we impose the invariance of the inner product
\ba\label{inv}
f^{abc}{}_e h^{ed}=-f^{abd}{}_e h^{ec}\,,
\ea
which is required when one will write down the Lagrangian in the future.
Unfortunately, Lagrangian of this nonabelian $(2,0)$ theory cannot be written down at this stage,
since the self-dual 2-form field $B_{\mu\nu,a}$ cannot be properly defined.
Although this is not the matter with our present discussion,
this must be a very important subject of future research.

\paragraph{Symmetry transformation}

The nonabelian $(2,0)$ theory is invariant under the gauge symmetry transformation
defined by
\ba\label{gauge}
&&\back
\delta_\Lambda X^I_a=\tilde \Lambda^b{}_a X^I_b\,,\quad
\delta_\Lambda \Psi_a=\tilde \Lambda^b{}_a \Psi_b\,,\quad
\delta_\Lambda H_{\mu\nu\rho,a}=\tilde \Lambda^b{}_a H_{\mu\nu\rho,b}\,,
\nt&&\back
\delta_\Lambda C^\mu_a=\tilde \Lambda^b{}_a C^\mu_b\,,\quad
\delta_\Lambda \tilde A_\mu{}^b{}_a=D_\mu \tilde \Lambda^b{}_a\,,
\ea
where
$\tilde A_\mu{}^b{}_a:=A_{\mu cd}f^{cdb}{}_a$
and $\tilde \Lambda^b{}_a:=\Lambda_{cd}f^{cdb}{}_a$.
%
And it is also invariant under the 6-dim $\cN=(2,0)$ supersymmetry transformation
\ba\label{SUSY}
\delta_\eps X_a^I \!&=&\! i\bar\eps\Gamma^I\Psi_a\nt
\delta_\eps \Psi_a \!&=&\!
 \Gamma^\mu\Gamma^ID_\mu X_a^I\eps
+\frac{1}{12}\Gamma_{\mu\nu\rho}H^{\mu\nu\rho}_a\eps
-\frac12 \Gamma_\mu \Gamma^{IJ}[C^\mu,X^I,X^J]_a\eps\nt
\delta_\eps H_{\mu\nu\rho,a} \!&=&\!
3i\bar\eps\Gamma_{[\mu\nu}D_{\rho]}\Psi_a
+i\bar\eps \Gamma^I\Gamma_{\mu\nu\rho\sig}[C^\sig,X^I,\Psi]_a\nt
\delta_\eps \tilde A_\mu{}^b{}_a \!&=&\!
 i\bar\eps \Gamma_{\mu\nu} C^\nu_c \Psi_d f^{cdb}{}_a \nt
\delta_\eps C^\mu_a \!&=&\! 0\,,
\ea
where $\eps$ and $\Psi$ are 32-component Majorana spinors
under the chirality condition
\ba\label{chiral}
\Gamma_{012345}\eps=+\eps\,,\quad
\Gamma_{012345}\Psi=-\Psi
\,.\ea
Thus 
the nonabelian $(2,0)$ theory 
is equipped with the expected symmetries of multiple M5-branes.
The main purpose of our work is 
to explore its properties through the reduction to branes in superstring theory 
and to clarify the availability of this formulation.
In the next section, starting from this theory, we will show that this theory actually reproduce the multiple D$p$-branes.

\section{D$p$-brane theory from nonabelian (2,0) theory}
\label{sec:Dp}

First we briefly review
how the nonabelian $(2,0)$ theory reproduces D4-brane action~\cite{Lambert:2010wm}.
In this case, we use the Lorentzian Lie 3-algebra $\{T^a,u_0,v^0\}$ defined by
\ba\label{Lor}
&&\back
[u_0,T^a,T^b]=if^{ab}{}_c T^c\,,\quad
[T^a,T^b,T^c]=-if^{abc}v^0\,,
\nt&&\back
\langle T^a,T^b \rangle=h^{ab}\,,\quad
\langle u_0,v^0 \rangle=1\,,\quad
{\rm otherwise}=0\,,
\ea
where $T^a$ are generators of the ordinary Lie algebra,
so this algebra is a central extension of Lie algebra.
Since $u_0-\alp v^0$ ($\alp>0$) is a negative norm generator,
the $u_0$- and $v^0$-component fields become ghosts.
Then we have to remove them in order to obtain a physical theory.
It is well known that this can be performed by the new kind of Higgs mechanism~\cite{Mukhi:2008ux,Ho:2008ei}.
In this mechanism, we assign a VEV (vacuum expectation value) to the $u_0$-component field without breaking gauge and supersymmetry.
When we set a VEV for the longitudinal field $C^\mu_{u_0}$,
D4-brane worldvolume theory can be reproduced
from the nonabelian $(2,0)$ theory.
In BLG theory, on the other hand,
we can obtain D2-brane worldvolume theory,
when we set a VEV for the transverse scalar field $X^I_{u_0}$.
In both cases,
the direction specified by the VEV becomes compactified
and then M-branes are reduced to D-branes in type IIA superstring theory.
In fact, the VEV can be interpreted as the compactification radius of the
M-theory direction.

In this section, we show that
the nonabelian $(2,0)$ theory can also reproduce D$p$-brane system ($p>4$)
on a torus $T^{p-4}$.
We realize this by using the central extension of Lorentzian Lie 3-algebra,
which is called the generalized loop algebra.
The number of its centers corresponds to the dimension of compactified torus.
It is already known that
BLG theory with this algebra reproduces D$p$-brane system
($p>2$) on a torus $T^{p-2}$~\cite{Ho:2009nk,Kobo:2009gz}. 
Therefore, the following discussion is similar to BLG theory case.

\subsection{Setup}

Now we start with the generalized loop algebra
$\{T^i_\vm,u_A,v^A\}$ \cite{Ho:2009nk,Kobo:2009gz} defined by
\ba\label{gla}
&&\back
{}[u_0,u_a,u_b]=0 
\nt&&\back
{}[u_0,u_a,T^i_\vm]=m_aT^i_\vm
\nt&&\back
{}[u_0,T^i_\vm,T^j_\vn]=m_av^a\delta_{\vm+\vn}\delta^{ij}
+if^{ij}{}_kT^k_{\vm+\vn}
\nt&&\back
{}[T^i_\vm,T^j_\vn,T^k_\vl]=
-if^{ijk}v^0\delta_{\vm+\vn+\vl}
\nt&&\back
\langle T^i_\vm,T^j_\vn\rangle=h^{ij}\delta_{\vm+\vn}\,,\quad
\langle u_A,v^B\rangle=\delta^B_A\,,\quad
{\rm otherwise}=0\,,
\ea
where $\vm,\vn,\vl\in \mathbb{Z}^d$, $A=0,1,\cdots,d$ and $a=1,\cdots,d$.
$f^{ij}{}_k$ ($i,j,k=1,\cdots,\dim{\mathfrak{g}}$) is a structure constant of an arbitrary Lie algebra $\mathfrak{g}$ defined as
\ba
{}[T^i,T^j]=if^{ij}{}_k T^k.
\ea
It can be easily shown that this Lie 3-algebra satisfies the fundamental identity (\ref{fund}) and
the invariant metric condition (\ref{inv}).
This algebra is characteristic in that 
the generators $u_A$ are not produced by any 3-brackets,
{\em i.e.}~$[\,\star,\star,\star\,]_{u_A}=0$,
and the generators $v^A$ are the center of the algebra,
{\em i.e.}~$[v^A,\star,\star\,]=0$.
According to systematic discussion in~\cite{Ho:2009nk},
these conditions are 
necessary if we want to remove ghost fields by the Higgs mechanism.

Actually, this algebra can be regarded as the original Lorentzian Lie 3-algebra (\ref{Lor}) with an infinite dimensional Lie algebra
$\{T^i_\vm,u_a,v^a\}$ given by
\ba\label{KM}
&&\back
{}[u_a,u_b]=0\,,\quad
{}[u_a,T^i_\vm]=m_aT^i_\vm\,,\quad
{}[T^i_\vm,T^j_\vn]=m_av^a\delta_{\vm+\vn}\delta^{ij}
+if^{ij}{}_kT^k_{\vm+\vn}\,,
\nt&&\back
\langle T^i_\vm,T^j_\vn\rangle=h^{ij}\delta_{\vm+\vn}\,,\quad
\langle u_a,v^b\rangle=\delta^b_a\,,\quad
{\rm otherwise}=0\,.
\ea
This is a higher loop generalization of the Kac-Moody algebra, and
can be regarded as a Lie algebra on a torus $T^d$.
As we mentioned, the nonabelian $(2,0)$ theory with Lorentzian Lie 3-algebra
reproduces D4-brane theory.
In our case, in the following discussion,
we define the higher dimensional fields
by collecting the infinite $T^i_\vm$-component fields and using Fourier transformation.
In other words, we interpret the index $\vm\in{\mathbb Z}^d$ as the Kaluza-Klein momentum along the torus $T^d$
to recover the higher dimension.
As a result, we will obtain the higher dimensional D$p$-brane theory
whose worldvolume is given by the flat torus $T^d$ bundle over the original
D4-brane worldvolume $\cM_5$ ({\em i.e.}~$p=4+d$).


\sss{Component Expansion}

Then, we expand all the fields into their components of Lie 3-algebra as
\ba
\Phi\!&=&\!\Phi_{(i\vm)}T^i_\vm+\Phi^A u_A+\udl{\Phi}_Av^A\nt
A_{\mu}\!&=&\!A_{\mu (i\vm)(j\vn)}T^i_\vm\wedge T^j_\vn
+\frac12 A^A_{\mu (i\vm)}u_A\wedge T^i_\vm
+A_{\mu}^{AB}u_A\wedge u_B+\cdots\,,
\ea
where $\Phi=X^I,\,\Psi,\,H_{\mu\nu\rho},\,C^\mu$. For simplicity, we set $A_{\mu}^{AB}=0$ in the following.
The omitted terms in the expansion of $A_{\mu}$ are the terms including $v^A$ which never appear in EOM's.

Each component of the covariant derivatives is written as
\ba
(D_\mu \Phi)_{(i\vm)}\!&=&\!
 (\hat D_\mu \Phi)_{(i\vm)}
+A'_{\mu(i\vm)}\Phi^0
+im_aA^0_{\mu(i\vm)}\Phi^a
\nt
(D_\mu \Phi)_{u_A} \!&=&\! \partial_\mu \Phi^A
\nt
(D_\mu \Phi)_{v^0}\!&=&\!
 \partial_\mu \udl\Phi_0
+im_a(A^a_{\mu(i\vm)}\Phi_{(i,-\vm)}+A_{\mu(i\vm)(i,-\vm)}\Phi^a)
\nt&&\!
-f^{ijk}A_{\mu(i\vm)(j\vn)}\Phi_{(k,-\vm-\vn)}
\nt
(D_\mu \Phi)_{v^a}\!&=&\!
 \partial_\mu \udl\Phi_a
-im_a(A^0_{\mu(i\vm)}\Phi_{(i,-\vm)}+A_{\mu(i\vm)(i,-\vm)}\Phi^0)\,,
\ea
where
\ba\label{ce}
(\hat D_\mu \Phi)_{(i\vm)}\!&=&\!
 \partial_\mu \Phi_{(i\vm)}
+f^{jk}{}_i A_{\mu(j,\vm-\vn)}^0\Phi_{(k\vn)}
\nt
A'_{\mu(i\vm)}\!&=&\!
-im_aA^a_{\mu(i\vm)}
+f^{jk}{}_iA_{\mu(j,\vm-\vn)(k\vn)}\,.
\ea


\sss{Solving the ghost sector}

The generalized loop algebra (\ref{gla}) has
$d+1$ negative norm generators $u_A-\alp v^A$ ($\alp>0$),
so the $u_A$ and $v^A$-component fields become ghosts.
Then one may wonder whether this theory is unitary. 
However, as we will see, 
it doesn't matter because these ghosts can be removed by the Higgs mechanism.
The detailed procedure is as follows.

First, we consider $u_A$-component fields.
Their EOM's are
\ba
\partial_\mu^2 X^{IA}=0\,,\quad
\Gamma^\mu \partial_\mu \Psi^A=0\,,\quad
\partial_{[\mu}H_{\nu\rho\sig]}^A=0\,,\quad
\partial_\mu C^{\nu A}=0\,.
\ea
The gauge transformation 
is given by
\ba
\delta_\Lambda X^{IA}=0\,,\quad
\delta_\Lambda \Psi^A=0\,,\quad
\delta_\Lambda H_{\mu\nu\rho}^A=0\,,\quad
\delta_\Lambda C^{\mu A}=0\,,
\ea
and the supersymmetry transformation is
\ba
&&\back
\delta_\eps X^{IA} = i\bar\eps\Gamma^I\Psi^A\,,\quad
\delta_\eps \Psi^A = \Gamma^\mu\Gamma^I\partial_\mu X^{IA}\eps\,,\quad
\delta_\eps H_{\mu\nu\rho}^A = 3i\bar\eps\Gamma_{[\mu\nu}\partial_{\rho]}\Psi^A\,,
\nt&&\back
\delta_\eps C^{\mu A}= 0\,.
\ea
This means that we can insert the VEV's as
\ba
X^{IA}={\rm const}.\,,\quad
\Psi^A=0\,,\quad
H_{\mu\nu\rho}^A={\rm arbitrary}\,,\quad
C^{\mu A}={\rm arbitrary}
\ea
without breaking gauge symmetry and supersymmetry. Then, in the following, we consider
\ba\label{VEV}
C^{\mu 0}=\lambda^0\delta^\mu_5\,,\quad
X^{Ia}=\lambda^{Ia}\,,\quad
{\rm otherwise}=0\,,
\ea
where $\vec\lam^a$ are constant vectors in ${\mathbb R}^5$
(the transverse directions of M5-branes), namely,
\ba
\vec\lam^a\in {\mathbb R}^d \subset {\mathbb R}^5.
\ea
In the following, we use $\{\vec\lam^a\}$ as the basis of ${\mathbb R}^d$.
Therefore, it is useful for later discussion
to define the dual basis $\vpi_a$ and the projection operator $P^{IJ}$ as
\ba\label{proj}
\vlam^a\cdot \vpi_b=\delta^a_b\,,\quad
P^{IJ}=\delta^{IJ}-\sum_a \lambda^{Ia}\pi^J_a\,.
\ea
The operator $P$ projects a vector onto subspace of ${\mathbb R}^5$ which is orthogonal to all $\vec\lam^a$, 
and it satisfies the projector condition $P^2=P$.
In the next subsection,
we will compactify this ${\mathbb R}^d$ space on a torus $T^d$,
and identify it with the torus $T^d$ defined by loop algebra (\ref{KM}).

Next, we look at $v^A$-component fields. 
For simplicity, we set $C^\mu_{(i\vm)}=0$ only here.
After setting VEV's (\ref{VEV}), their EOM's become
\ba
0\!&=&\!
D^2_\mu \udl X^I_0
\nt&=&\!
D^2_\mu \udl X^I_a
+\frac{1}{2}m_a\lambda^0\bar\Psi_{(i\vm)}\hat\Gamma^I \Psi_{(i,-\vm)}
\nt&&\!\qquad\quad
-m_a^2(\lam^0)^2\lam^{[Ia} X^{J]}_{(i\vm)} X^J_{(i,-\vm)}
-m_a(\lam^0)^2f^{ijk} X^J_{(i\vm)}X^J_{(j\vn)}X^I_{(k,-\vm-\vn)}
\nt
0\!&=&\!
\Gamma^\mu D_\mu\udl\Psi_0
=
\Gamma^\mu D_\mu\udl\Psi_a
-im_a\lambda^0 X^I_{(i\vm)} \hat\Gamma^I \Psi_{(i,-\vm)}
\nt
0\!&=&\!
D_{[\mu}\udl{H}_{\nu\rho\sig],0}
=
D_{[\mu}\udl{H}_{\nu\rho\sig],a}
+\eps_{\mu\nu\rho\sig 5\tau}m_a\lambda^0\left(
  \frac{1}{4} X^I_{(i\vm)} D^\tau X^I_{(i,-\vm)}
 +\frac{i}{8} \bar\Psi_{(i\vm)} \Gamma^\tau \Psi_{(i,-\vm)}
\right)
\nt
0\!&=&\!
\udl{\tilde F}_{\mu\nu}{}^a{}_{0}
=
\udl{\tilde F}_{\mu\nu}{}^{(i\vm)}{}_{0}
=
\udl{\tilde F}_{\mu\nu}{}^{0}{}_{a}
=
\udl{\tilde F}_{\mu\nu}{}^{(i\vm)}{}_{a}
 -m_a\lambda^0 H_{\mu\nu 5(i\vm)}
\nt
0\!&=&\!
D_\mu \udl C^\nu_A
\,,
\ea
where $\hat \Gamma^I:=i\Gamma_5\Gamma^I$ and these satisfies $\frac12\{\hat\Gamma^I,\hat\Gamma^J\}=\delta^{IJ}$.\footnote{
In our notation,
$\frac12\{\Gamma_\mu,\Gamma_\nu\}=g_{\mu\nu}={\rm diag.}\,(-+\cdots+)$
and $\frac12\{\Gamma^I,\Gamma^J\}=\delta^{IJ}$.}
Note that all the equations of $v^0$-component fields are free,
while the equations of $v^a$-component fields are necessarily not.
This doesn't matter as long as we consider the VEV's of $u_A$-component fields to be constants.

\subsection{Derivation of D$p$-brane action}

Now we concentrate on the EOM's for $T^i_\vm$-component fields.
In order to obtain the D$p$-brane action,
we compactify the ${\mathbb R}^d$ space spanned by $\vec\lam^a$ on a torus $T^d$ and
regard the index $\vm\in \mathbb{Z}^d$ as the Kaluza-Klein momentum along the torus.
Then we identify the infinite $T^i_\vm$-component fields with the $(6+d)$-dim fields through the Fourier transformation on it:
\ba\label{redef}
\hat\Phi_i(x,y):=\sum_{\vm} \Phi_{(i\vm)}(x)e^{-i\vm\cdot\vy}\,,\quad
\hat A_{\mu i}(x,y):=\sum_{\vm} A^0_{\mu(i\vm)}(x)e^{-i\vm\cdot\vy}\,,
\ea
where $x^\mu$ are coordinates of M5-brane worldvolume,
and $y^a\in [0,2\pi]$ are coordinates of the $d$-dim torus $T^d$~\cite{Ho:2009nk,Kobo:2009gz}.
We will also use the notation of field strength
\ba\label{redef-F}
\hat F_{\mu\nu,i}(x,y):=\sum_{\vm} F^0_{\mu\nu(i\vm)}(x)e^{-i\vm\cdot\vy}\,,
\ea
where
$F^0_{\mu\nu(i\vm)}
:=\partial_\mu A^0_{\nu(i\vm)}-\partial_\nu A^0_{\mu(i\vm)}
+f^{jk}{}_{i}A^0_{\mu(j,\vm-\vn)}A^0_{\nu(k\vn)}$\,.
In fact, this procedure corresponds to taking the field theoretical T-duality~\cite{Taylor:1996ik}
for the directions of $T^d$,
since it means that we make the brane worldvolume extended to these directions.

\paragraph{$C^\mu$-field and constraints}
After inserting the VEV's (\ref{VEV}), the EOM (\ref{EOM}) for $C^\mu$-field and the constraints (\ref{cons}) become
\ba\label{cons-p}
D_5 X^I_{(i\vm)}
=D_5 \Psi_{(i\vm)}
=D_5 H_{\mu\rho\sig(i\vm)}
=D_\mu C^\nu_{(i\vm)}
=C^\alp_{(i\vm)}
=0\,,
\ea
where $\alp=0,\cdots,4$.
Also, from eq.\,(\ref{gauge}) and (\ref{SUSY}),
we find that one can set a VEV as
\ba\label{C}
C^5_{(i\vm)}={\rm const}.
\ea
without breaking gauge symmetry and supersymmetry.
However, as we will see, 
this field and its VEV has no influence on the EOM's in the final form.

\paragraph{Spinor field}

After inserting the VEV's (\ref{VEV}), we obtain
\ba
{}[C^\mu,X^I,\Psi]_{(i\vm)}
= \lambda^0\delta^\mu_5\left(
 m_a\lambda^{Ia}\Psi_{(i\vm)}
+if^{jk}{}_iX^I_{(j\vn)}\Psi_{(k,\vm-\vn)}
\right)\,.
\ea
Then, using the projector (\ref{proj}), we define the field $A_{a(i\vm)}$ as
\ba\label{X}
X^I_{(i\vm)}=P^{IJ}X^J_{(i\vm)} + \lambda^{Ia}(\vpi_a\cdot \vec X)_{(i\vm)}
=:P^{IJ}X^J_{(i\vm)} + \lambda^{Ia}A_{a(i\vm)}\,.
\ea
This field can be regarded as the gauge field along the fiber torus $T^d$.
Therefore, by using these equations and eq.\,(\ref{cons-p}),
the EOM (\ref{EOM}) for spinor field becomes
\ba
0\!&=&\!
\Gamma^\alp \hat D_\alp\Psi_{(i\vm)}
+\lambda^0\lambda^{Ia}\Gamma_5\Gamma^I
(m_a\Psi_{(i\vm)}+if^{jk}{}_iA_{a(j\vn)}\Psi_{(k,\vm-\vn)})
\nt&&\!
+\lambda^0\Gamma_5\Gamma^I[P^{IJ}X^J,\Psi]_{(i\vm)}\,.
\ea
After the field redefinition (\ref{redef}), this can be represented as
\ba\label{Psi-p}
0\!&=&\!
\Gamma^\alp \hat D_\alp\hat\Psi
+\Gamma^a \hat D_a \hat\Psi
+\lambda^0\hat\Gamma^I[P^{IJ}\hat X^J,\hat \Psi]\,,
\ea
where the covariant derivative is defined as
$
\hat D_a\hat\Phi_i
:=\partial_a\hat\Phi_i-i[\hat A_{a},\hat\Phi]
$.
The $\Gamma$-matrices
$\Gamma^a:=i\lambda^0\lambda^{Ia}\Gamma_5\Gamma^I$
satisfy
$\frac12\{\Gamma^a,\Gamma^b\}=g^{ab}$ which is the metric on the torus $T^d$ given by
\ba\label{metric-T}
g^{ab}:=|\vlam^0|^2\vlam^a\cdot\vlam^b\,.
\ea

\paragraph{Scalar fields}

Similarly, after inserting the VEV's, we obtain
\ba
{}[C^\mu,X^I,X^J]_{(i\vm)}
=
\lambda^0\delta^\mu_5 \left(
m_a\lambda^{[Ia}X^{J]}_{(i\vm)}
+if^{jk}{}_iX^{[I}_{(j\vn)}X^{J]}_{(k,\vm-\vn)}
\right)\,.
\ea
Then, by using eq.\,(\ref{cons-p}) and (\ref{X}),
we obtain
\ba
(D_\alp^2 X^I)_{(i\vm)}
=
 P^{IJ}(\hat D_\alp^2 X^J)_{(i\vm)}
+\lambda^{Ia}(\hat D^\alp F_{\alp a})_{(i\vm)}\,,
\ea
where
$(F_{\alp a})_{(i\vm)}:=\hat D_\alp A_{a(i\vm)}+im_aA_{\alp(i\vm)}^0$\,.
After the field redefinition,
the EOM's (\ref{EOM}) for scalar fields become
\ba\label{X-p}
0\!&=&\!
 P^{IJ}\hat D_\alp^2 \hat X^J 
+P^{IJ}\hat D_a^2 \hat X^J 
\nt&&\!
+i(\lambda^0)^2\lambda^{Ia}[P^{JL}\hat X^L,P^{JK}\hat D_a\hat X^K]
-(\lambda^0)^2[P^{JM}\hat X^M,[P^{JL}\hat X^L,P^{IK}\hat X^K]]
\nt&&\!
+\lambda^{Ib}(\hat D^\alp \hat F_{\alp b})
+\lambda^{Ib}(\hat D^a \hat F_{ab})
+\frac{i\lambda^0}{2}[\hat{\bar\Psi},\hat\Gamma^I\hat\Psi]\,,
\ea
where $\hat D^a=g^{ab}\hat D_b$
and $\hat F_{ab}:=\partial_a \hat A_b -\partial_b \hat A_a -i[\hat A_a,\hat A_b]$\,.

\paragraph{Gauge field}

The EOM for gauge field becomes
\ba\label{F-p}
0\!&=&\!
\tilde F_{\mu\nu}{}^{(j\vn)}{}_{(i\vm)}
 -i\lambda^0f^{kj}{}_i H_{\mu\nu 5(k,\vm-\vn)}
\nt&=&\!
\tilde F_{\mu\nu}{}^0{}_{(i\vm)}
 -if^{jk}{}_i C^5_{(j\vn)} H_{\mu\nu 5(k,\vm-\vn)}
\nt&=&\!
\tilde F_{\mu\nu}{}^a{}_{(i\vm)}
 +m_a\lambda^0 H_{\mu\nu 5(i\vm)}\,.
\ea
In fact, we don't use the second equation in the following,
since we now regard only $A_{\mu(i\vm)}^0$ as the gauge field, 
as we can see in eq.\,(\ref{ce}) or (\ref{redef}).
This is a direct reason why $C^5_{(im)}$-field gives no effects on the EOM's in the final form. 

\paragraph{2-form field}

Similarly, 
the EOM for self-dual 2-form field becomes
\ba
0\!&=&\!
\hat D_{[\mu}H_{\nu\rho\sig](i\vm)}
+\frac{\lambda^0}4\eps_{\mu\nu\rho\sig5\tau}
 [P^{IJ}X^J,P^{IK}\hat D^\tau X^K]_{(i\vm)}
+\frac{\lambda^0\lambda^{Ia}}4\eps_{\mu\nu\rho\sig5\tau}
 P^{IJ}\hat D^\tau \hat D_a X^J_{(i\vm)}
\nt&&\!
+\frac{1}{\lambda^0}\eps_{\mu\nu\rho\sig 5\tau}\hat D^a F_{a\tau(i\vm)}
+\frac{i\lambda^0}{8}\eps_{\mu\nu\rho\sig 5\tau}[\bar\Psi,\Gamma^\tau\Psi]_{(i\vm)}\,.
\ea
Then, by using eq.\,(\ref{F-p}), the self-duality of $H_{\mu\nu\rho}$ (\ref{dual}),
and the field redefinition (\ref{redef-F}), this can be rewritten as
\ba\label{H-p}
0
\!&=&\!
 \frac{1}{(\lambda^0)^2}\left(
 \hat D^\alp \hat F_{\alp\beta}
 +\hat D^a \hat F_{a\beta}\right)
+i [P^{IJ}\hat X^J,P^{IK}D_\beta \hat X^K]
-\frac{1}{2} [\hat{\bar\Psi},\Gamma_\beta\hat\Psi]\,.
\ea

\paragraph{Summary}

First, we note that
the Higgs mechanism removes the ghost sector completely
without breaking gauge symmetry and supersymmetry.
In fact, the ghost fields {\em never} appear in the EOM's for $T^i_\vm$-component fields.

Then we can finally show that all the EOM's derived above,
{\em i.e.}~eq.\,(\ref{cons-p}), (\ref{Psi-p}), (\ref{X-p}), (\ref{F-p}) and (\ref{H-p}),
are successfully reproduced
from the $(5+d)$-dim super Yang-Mills action
\ba\label{Lag}
S\!&=&\!\lam^0\int d^5x\,\frac{d^dy}{(2\pi)^d}\,\sqrt{g}\,\cL\,,
\nt
\cL
\!&=&\!
-\frac12(\hat D_\umu \hat X^I)P^{IJ}(\hat D^\umu \hat X^J)
+\frac{i}{2}\hat{\bar\Psi}\Gamma^\umu\hat D_\umu \hat \Psi
-\frac{1}{4(\lambda^0)^2}\hat F_{\umu\unu}^2
\nt&&\!
-\frac{(\lambda^0)^2}{4}[P^{IK}\hat X^K,P^{JL}\hat X^L]^2
+\frac{i\lambda^0}{2}\hat{\bar\Psi}\hat\Gamma^I[P^{IJ}\hat X^J,\hat\Psi]\,.
\ea
where the spacetime indices are summarized as $\umu=(\alpha,a)$,
and $g:=\det\,g^{ab}$.
This is nothing but the low energy effective action of multiple D$p$-branes ($p=4+d$) on $\cM_5\times T^d$.
Therefore, we conclude that one can reproduce D$p$-brane system from
nonabelian $(2,0)$ theory. 

\section{NS5-brane theory from nonabelian (2,0) theory}
\label{sec:NS5}

In the previous section, we successfully derive D$p$-brane system on a torus $T^{p-4}$
from the nonabelian $(2,0)$ theory by using the Higgs mechanism (\ref{VEV}) and
the field redefinition (\ref{redef}). 
Let us see here the physical meaning of each step.
From the discussion in Lorentzian BLG theory,
it is well known that  
putting a VEV of $u_A$-component field corresponds to the compactification.
Therefore, in eq.\,(\ref{VEV}),
we put a VEV $C^{\mu 0}$ to compactify one of the $x^\mu$-directions which becomes M-theory direction,
and then we also put VEV's $X^{I a}$ to compactify some of the $x^I$-directions.
After the field redefinition (\ref{redef}) which is equivalent to
the field theoretical T-duality for the latter compactified directions, 
we finally obtain D$p$-brane system on a torus $T^{p-4}$.

In this section, we change the way of setting VEV's from the previous case.
This should correspond to changing the directions of M-compactification
and that of taking T-duality.
Especially, we now consider the reduction to type IIA/IIB NS5-brane system,
and investigate whether these branes can be reproduced from 
the nonabelian (2,0) theory. 

\subsection{Type IIA NS5-brane theory}
\label{sec:NS5A}

In order to obtain type IIA NS5-branes from M5-branes,
we change the direction of M-compactification,
compared with D4-brane case.
Therefore, here we use the original Lorentzian Lie 3-algebra $\{T^a,u_0,v^0\}$ defined by
\ba
&&\back
[u_0,T^a,T^b]=if^{ab}{}_c T^c\,,\quad
[T^a,T^b,T^c]=-if^{abc}v^0\,,
\nt&&\back
\langle T^a,T^b \rangle=h^{ab}\,,\quad
\langle u_0,v^0 \rangle=1\,,\quad
{\rm otherwise}=0\,.
\ea
In D4-brane case, we put a non-zero VEV into the longitudinal field $C^{\mu 0}$
in order to compactify one of $x^\mu$-direction.
Then in this case, we put a VEV into $u_0$-components as
\ba
X^{I0}=\lambda \delta^I_{10}\,,\quad
{\rm otherwise}=0\,,
\ea
in order to compactify one of the transverse $x^I$-direction as M-theory direction.

\paragraph{On gauge field}

In this setup, the EOM for gauge field $\tilde A_\mu{}^b{}_a$ is
\ba
\tilde F_{\mu\nu}{}^b{}_a 
= 0\,,
\ea
and its supersymmetry transformation is
\ba
\delta_\eps \tilde A_\mu{}^b{}_a = 0\,.
\ea
This means that the gauge field $\tilde A_\mu{}^b{}_a$ 
have no physical degrees of freedom,
and can be set to zero up to gauge transformation.
Therefore, the covariant derivative $\hat D_\mu$ in eq.\,(\ref{ce}) 
is reduced to the partial derivative $\partial_\mu$.

\paragraph{Equations of motion}
The remaining EOM's are
\ba\label{EOM-NS5A}
\partial_\mu^2X_a^i
-\lambda^2[C^\mu,[C_\mu,X^i]]_a
\!&=&\!0\nt
\partial_\mu^2X_a^{10}
\!&=&\!0\nt
\Gamma^\mu \partial_\mu\Psi_a-\lambda\Gamma_\mu\Gamma^{10}[C^\mu,\Psi]_a
\!&=&\!0\nt
\partial_{[\mu}H_{\nu\rho\sig]a}
-\frac{\lambda}4\eps_{\mu\nu\rho\sig\lam\tau}[C^\lam,(\partial^\tau X^{10}+\lam\tilde A^{\tau 0})]_a
\!&=&\!0\nt
\tilde F_{\mu\nu}{}^0{}_a-[C^{\rho},H_{\mu\nu\rho}]_a
\!&=&\!0\nt
\partial_\mu C^\nu_a
\!&=&\!0
\ea
where $i=6,\cdots,9$,
and we set $\partial^\mu \tilde A_{\mu}{}^0{}_a=0$ using the gauge transformation.

For the multiple D$p$-branes, the interaction terms like $[X,[X,X]]$ or $[X,\Psi]$
come from strings ending on different branes.
In this case, however, $C^{\mu}$-field has no dynamical degrees of freedom because they have no kinetic terms.
Therefore, we naively guess that the terms including this field doesn't describe 
the interaction between different NS5-branes,
and so the resultant EOM's (\ref{EOM-NS5A}) seem practically
the simple copies of free theory of $\cN=(2,0)$ multiplet.
In order to obtain the interaction terms, 
we need to go beyond the present construction of the nonabelian (2,0) theory. 

\subsection{Type IIB NS5-brane theory}
\label{sec:NS5B}

In order to obtain type IIB NS5-branes from M5-branes,
we interchange the direction of M-compactification and that of taking T-duality,
compared with D5-brane case.
Therefore, in this case,
we use a generalized loop algebra $\{T^i_m,u_{0,1},v^{0,1}\}$ defined by
\ba\label{loop-B}
&&\back
[u_0,u_1,T^i_m]=mT^i_m\,,\quad
[u_0,T^i_m,T^j_n]=mv^1\delta_{m+n}\delta^{ij}+if^{ij}{}_kT^k_{m+n}\,,
\nt&&\back
[T^i_m,T^j_n,T^k_l]=-if^{ijk}v^0\delta_{m+n+l}\,,
\nt&&\back
\langle T^i_m,T^j_n \rangle=h^{ij}\delta_{m+n}\,,\quad
\langle u_0,v^0 \rangle
=\langle u_1,v^1 \rangle=1\,,\quad
{\rm otherwise}=0\,.
\ea
In D5-brane case, we put non-zero VEV's into $C^{\mu 0}$ and $X^{I1}$
as eq.\,(\ref{VEV}). 
Then, we now put VEV's into $u_{0,1}$-components as
\ba\label{VEV-B}
X^{I0}=\lam^0 \delta^I_{10}\,,\quad
C^{\mu 1}=\lam^1 \delta^\mu_5\,,\quad
{\rm otherwise}=0\,.
\ea
We also redefine the fields in a similar but slightly different way from eq.\,(\ref{redef}) as
\ba\label{redef-B}
\hat \Phi_i(x,y)=\sum_m \Phi_{(im)}(x)e^{-imy}\,,\quad
\hat A_{\mu,i}(x,y)=\sum_m A^1_{\mu(im)}(x)e^{-imy}\,,\quad
\cdots\,.
\ea
Note that we now regard $A^1_{\mu(im)}$ field as the gauge field,
while we use $A^0_{\mu(im)}$ field in D5-brane case (\ref{redef}).

\paragraph{$C$-field and constraints}
The EOM for $C$-field and the constraints become
\ba\label{cons-B}
D_5 X^I_{(im)}
=D_5 \Psi_{(im)}
=D_5 H_{\nu\rho\sig(im)}
=D_\mu C^\nu_{(im)}
=C^\alp_{(im)}
=0\,.
\ea
where $\mu=0,\cdots,5$ and $\alp=0,\cdots,4$.

\paragraph{Gauge field}

The EOM for gauge field becomes
\ba\label{FH-B2}
\tilde F_{\mu\nu}{}^0{}_{(im)}-m\lam^1 H_{\mu\nu 5(im)}
-if^{jk}{}_i C^5_{(j,m-n)}H_{\mu\nu 5(kn)}
\!&=&\!0\nt
\tilde F_{\mu\nu}{}^1{}_{(im)}
=\tilde F_{\mu\nu}{}^{(jn)}{}_{(im)}
\!&=&\!0\,,
\ea
and the supersymmetry transformation becomes
\ba
&&\back
\delta_\eps \tilde A_\mu{}^0{}_{(im)}
=i\bar\eps\Gamma_{\mu 5}\left(m\lam^1\Psi_{(im)}+if^{jk}{}_i C^5_{(j,m-n)}\Psi_{(kn)}\right)
\nt&&\back
\delta_\eps \tilde A_\mu{}^1{}_{(im)}
=\delta_\eps \tilde A_\mu{}^{(jn)}{}_{(im)}
=0\,.
\ea
Therefore, we can see that
$\tilde A_\mu{}^1{}_{(im)}$ and $\tilde A_\mu{}^{(jn)}{}_{(im)}$
have no physical degrees of freedom, and can be set to zero up to
gauge transformation.
This means that the covariant derivative
$\hat D_\alp\Phi_{(im)}
=\partial_\alp\Phi_{(im)}
-i\tilde A_\mu{}^{(jn)}{}_{(im)}\Phi_{(jn)}$
is reduced to the partial derivative.
Moreover, $\tilde F_{\mu\nu}{}^0{}_{(im)}$ is also reduced to
\ba
\tilde F_{\mu\nu}{}^0{}_{(im)}
\!&=&\!
 \partial_\mu \tilde A_\nu{}^0{}_{(im)}
-\partial_\nu \tilde A_\mu{}^0{}_{(im)}
\nt&=&\!
 m\bigl(\partial_\mu A^1_{\nu(im)}-\partial_\nu A^1_{\mu(im)}\bigr)
+if^{jk}{}_i\bigl(\partial_\mu A_{\nu(j,m-n)(kn)}-\partial_\nu A_{\mu(j,m-n)(kn)}\bigr)
\,.~~
\ea
Then from eq.\,(\ref{FH-B2}), we obtain
\ba\label{FH-B}
F^1_{\mu\nu(im)}
\!&:=&\!
\partial_\mu A^1_{\nu(im)}-\partial_\nu A^1_{\mu(im)}
=\lam^1 H_{\mu\nu 5(im)}
\nt
F_{\mu\nu(im)(jn)}
\!&:=&\!
\partial_\mu A_{\nu(im)(jn)}-\partial_\nu A_{\mu(im)(jn)}
=C^5_{(im)} H_{\mu\nu 5(jn)}
\,.
\ea
Here we define the field strength $F_{\mu\nu}$,
but unfortunately, the interaction term like\\
$f^{jk}{}_i A^1_{\mu(j,m-n)} A^1_{\nu(kn)}$
cannot appear in this setup.

\paragraph{Scalar and spinor fields}
Then, the EOM's for scalar fields and spinor fields are
\ba\label{X-B}
\hat D_\alp^2\hat X^i
+\hat D_y^2\hat X^i
\!&=&\!0\nt
\Gamma^\alp \hat D_\alp \hat\Psi
+\Gamma^y \hat D_y \hat\Psi
\!&=&\!0
\ea
where $i=6,\cdots,9$, and we define
\ba
\hat D_y \hat\Phi := \partial_y \hat\Phi - i[\hat C_y, \hat \Phi]\,,\quad
\hat C_y:=-\frac{1}{\lam^1}\hat C_5\,,\quad
\Gamma^y:=i\lam^0\lam^1\Gamma_5\Gamma^{10}\,,
\ea
satisfying $\frac12\{\Gamma^y,\Gamma^y\}=g^{yy}=(\lambda^0\lambda^1)^2$.
Note that $\hat C_y$-field has no kinetic terms,
so it is not a gauge field,
although the theory in this setup is invariant under the transformation
\ba
\delta_\Lambda \Phi_{(im)}=if^{jk}{}_{i}\Lambda_{(j,m-n)}\Phi_{(kn)}\,,\quad
\delta_\Lambda C_{y(im)}=\hat D_y\Lambda_{(im)}\,.
\ea
This means that the covariant derivative $\hat D_y$ can be also reduced to
the partial derivative if we gauge away the $\hat C_y$-field.
Anyway, it is interesting that $C^\mu$-field appears in EOM's, 
which is different from 
D5-brane case.

The remaining EOM for the scalar field is
\ba\label{X10}
\hat D^\alp \bigl(\hat D_\alp X^{10}_{(im)}+\lam^0 A'_{\alp(im)}\bigr)
=0\,,
\ea
where
$A'_{\alp(im)}$ is defined in eq.\,(\ref{ce}).
Here, by using eq.\,(\ref{redef-B}) and (\ref{FH-B}), we can see that
\ba\label{DA}
\hat D_y \hat A_{\alp,i}=\sum_m A'_{\alp(im)}e^{-imy}
\ea
is satisfied. Therefore, if we redefine the field as
\ba\label{Ay}
\hat A_y:=-\frac{1}{\lam^0}\hat X^{10}\,,
\ea
we can define the field strength $\hat F_{\alp y}$ and show that
\ba\label{F-B}
\hat D^\alp \hat F_{\alp y}
\!&:=&\!
\hat D^\alp\bigl(
 \hat D_\alp \hat A_y-\hat D_y \hat A_\alp-i[\hat A_{\alp},\hat A_y]
\bigr)
=
-i[\hat D^\alp\hat A_{\alp},\hat A_y]
=0\,,
\ea
where we use eq.\,(\ref{X10}) at the second equality,
and the last equality is satisfied up to gauge transformation.

\paragraph{2-form field}

Using the above results, the EOM for 2-form field
\ba
\hat D_{[\mu}\hat H_{\nu\rho\sig]}
-\frac{i\lam^0\lam^1}4\eps_{\mu\nu\rho\sig 5\tau}
 \hat D_y (\hat D_\tau \hat X^{10} + \lam^0\hat A'_\tau)
\!&=&\!0
\ea
can be rewritten, by using eq.\,(\ref{FH-B}) for the first term
and eq.\,(\ref{X10})--(\ref{Ay}) for the second term, as
\ba\label{H-B}
\hat D^\beta \hat F_{\alp\beta}+\hat D^y \hat F_{\alp y}
\!&=&\!0\,,
\ea
where we use $\hat D^y [\hat A_\alp,\hat A_y]=0$ up to gauge transformation,
similarly to eq.\,(\ref{F-B}).

\paragraph{Summary}

We have obtained all the EOM's (\ref{cons-B}), (\ref{FH-B}), (\ref{X-B}),
(\ref{F-B}) and (\ref{H-B}).
Note that they are practically free part of the EOM's of 6-dim $\cN=(1,1)$ super Yang-Mills theory
which is known as the low energy effective theory of type IIB NS5-branes.
Therefore, we conclude that one can partially reproduce the type IIB NS5-brane theory on $\cM_5\times S^1$
from the nonabelian $(2,0)$ theory. Further justification from the viewpoint of S-duality will be done in \S\,\ref{sec:U-D5}.

Finally, 
let us look at the kinetic part of the theory.
The EOM's of original nonabelian $(2,0)$ theory can be reproduced from
the Lagrangian
\ba
\cL
\!&=&\!
-\frac12(D_\mu X^I)^2
+\frac{i}{2}\bar\Psi\Gamma^\mu D_\mu \Psi
-\frac{1}{12}H_{\mu\nu\rho}^2
%
+\cdots\,.
\ea
Then, by using the field redefinition (\ref{redef-B}) and (\ref{FH-B}),
this Lagrangian becomes
\ba\label{Lag-B}
\cL
\!&=&\!
-\frac12(\hat D_\umu \hat X^i)^2
+\frac{i}{2}\hat{\bar\Psi}\Gamma^\umu \hat D_\umu \hat\Psi
-\frac{1}{4(\lam^1)^2}\hat F_{\umu\unu}^2
+\cdots\,,
\ea
where $\umu=(\alp,y)$.
This is nothing but the kinetic part of 6-dim $\cN=(1,1)$ super
Yang-Mills Lagrangian.
However, we should remind that
$\hat D_\umu$ is {\em not} the covariant derivative,
that is, it does {\em not} include the gauge field $\hat A_\umu$:
In fact, both $\hat D_\alp$ and $\hat D_y$ are simply the partial derivatives up to gauge transformation.
In order to make $\hat D_\umu$ the covariant derivative and also to obtain
all the interaction terms 
in super Yang-Mills Lagrangian, we must generalize the original nonabelian $(2,0)$ theory.
This must be a very interesting subject,
but we put off detailed discussion as a future work.

\section{More comments on nonabelian $(2,0)$ theory}
\label{sec:gen}

\subsection{Generalization of setting VEV's and total derivative terms}
\label{sec:genV}

In the previous sections, 
we chose the VEV's as eq.\,(\ref{VEV}) for D$p$-branes
or as eq.\,(\ref{VEV-B}) for type IIB NS5-branes. 
This means that we have seen only the case
where the direction of M-compactification and that of taking T-duality are
perpendicular to each other.


If we want to discuss more general cases where the directions are
not perpendicular, we may turn on an additional VEV $C^{\mu a}$
or $X^{I1}$ as
\ba\label{VEV2}
&&\back
C^{\mu 0}=\lambda^0\delta^\mu_5\,,\quad~
C^{\mu a}=\tilde\lambda^a\delta^\mu_5\,,\quad~
X^{Ia}=\lambda^{Ia}\qquad~\text{for D$p$-branes}
\nt&&\back
X^{I0}=\lambda^{0}\delta^I_{10}\,,\quad
X^{I1}=\tilde\lambda^{1}\delta^I_{10}\,,\quad
C^{\mu 1}=\lambda^1\delta^\mu_5\qquad \text{for type IIB NS5-branes}
\ea
since putting these VEV's can be regarded as
the M-compactification for the direction of
\ba\label{lam0}
&&\back
\vlam^0=(\vec 0,\lambda^0;0,0,0,0,0,0)
\qquad~~~\text{for D$p$-branes}
\nt&&\back
\vlam^0=(\vec 0,0\,;0,0,0,0,0,\lam^0l_p^3)
\qquad\text{for type IIB NS5-branes}
\ea
and taking T-duality for the direction of
\ba\label{lama}
&&\back
\vlam^a=(\vec 0,\tilde\lambda^a;
\lambda^{Ia}l_p^3)
\qquad\qquad\qquad\,\,\text{for D$p$-branes}
\nt&&\back
\vlam^1=(\vec 0,\lam^1;0,0,0,0,0,\tilde\lam^1l_p^3)
\qquad\text{for type IIB NS5-branes}
\ea
where $\vec 0$ is the $(4+1)$-dim zero vector, and $l_p$ is 11-dim Planck length.
Note that we now recover the factors $l_p^3$
which were previously set to 1.
They have to appear here, since the canonical mass dimension of $C^\mu$ (and $\vec\lam^{0,a}$) is $-1$,
while that of $X^I$ is $2$.

After a straightforward calculation, we can show that
this generalization of setting VEV's (\ref{VEV2}) does {\em not}
change any terms of the EOM's in all the cases.
This means that this generalization affects at most only the terms
which doesn't appear in EOM's, for example, total derivative terms in Lagrangian.
In fact, it is well known that
such a shift of T-duality directions 
corresponds to T-transformation which affects
the Chern-Simons term in D$p$-brane Lagrangian.
To see this, therefore, we now try to discuss total derivative terms in Lagrangian
of the nonabelian $(2,0)$ theory.

Since the nonabelian (2,0) theory must not have dimensionful parameters,
we only consider the total derivative terms with mass dimension 6.
Then one natural candidate is 
\ba\label{td}
\cL~\supset~
\eps^{\mu\nu\rho\sig\lam\tau}\tilde F_{\mu\nu}{}^a{}_b
\tilde F_{\rho\sig}{}^b{}_c \tilde F_{\lam\tau}{}^c{}_a\,.
\ea
Let us now consider the D$p$-brane $(p>4)$ case with VEV's (\ref{VEV2}).
In this case,
both $\vlam^0$ and $\vlam^a$ have nonzero elements for $x^5$-direction,
so the projector (\ref{proj}) must be redefined as
\ba
P^{MN}=\delta^{MN}-\sum_A \lam^{MA}\pi^N_A\,,\quad
\vec\lam^A\cdot\vec\pi^B=\delta^A_B\,,
\ea
where $M,N=5,6,\cdots,10$ and $A=0,1,\cdots, d\,(=p-4)$.
By using this, the gauge field $A_{a(i\vm)}$ can be defined like as eq.\,(\ref{X})
\ba
X^M_{(i\vm)}
\!&=&\!
P^{MN}X^N_{(i\vm)}+\lambda^{MA}(\vec\pi_A\cdot\vec X)_{(i\vm)}
\nt&=:&\!
P^{MN}X^N_{(i\vm)}+\lambda^{M0}(\vec\pi_0\cdot\vec X)_{(i\vm)}
+\lam^{Ma}A_{a(i\vm)}\,,
\ea
where we naturally define as
\ba\label{X5}
X^5_{(i\vm)}:=\frac{1}{\lam^0}A^0_{\mu=5,(i\vm)}\,,\quad
X^5_{u_A}:=C^{5A}\,.
\ea
Note that we set $l_p=1$ again for readability.
Therefore, the nontrivial factor in eq.\,(\ref{td}) can be written as
\ba
F^0_{\mu 5,(i\vm)}
\!&=&\!
\lam^0 D_\mu X^5_{(i\vm)}-\partial_5 A^0_{\mu(i\vm)}
\nt&=&\!
\lam^0 \left[
 P^{5M} \hat D_\mu X^M_{(i\vm)}
 + \lam^0 F_{\mu 0(i\vm)}
 + \sum_a \tilde\lam^{a} F_{\mu a(i\vm)} \right]
-\partial_5 A^0_{\mu(i\vm)}\,,
\ea
where
$
F_{\mu 0(i\vec m)}
:= \hat D_\mu (\vec \pi_0\cdot\vec X)_{(i\vec m)}
+ A'_{\mu (i\vec m)}
$ and
$
F_{\mu a(i\vec m)}
:= \hat D_\mu A_{a(i\vm)} 
+ i m_a A^0_{\mu (i\vec m)}
$.
The notation of other fields is defined around eq.\,(\ref{ce}) and (\ref{redef-F}).
Then we obtain the total derivative terms in D$p$-brane action
which can be derived from the term (\ref{td}) as
\ba\label{Lag-CS}
S ~\supset~ \int d^5x \,\frac{d^dy}{(2\pi)^d}\,\sqrt{g}
\left[(\lam^0)^2\tilde\lam^a
\epsilon^{\mu\nu\rho\sig\lam 5}
\hat F_{\mu\nu,i}\hat F_{\rho\sig,j}\hat F_{\lam a,k}
f^{il}{}_m f^{jm}{}_n f^{kn}{}_l
+\cdots\right]\,,
\ea
where `$\cdots$' are the total derivative terms which don't vanish in the $\tilde\lam^a\to 0$ limit.
We neglect them here,
since it is known that
the total derivative terms don't play any role,
when M-compactification direction is perpendicular to T-duality direction,
{\em i.e.}~$\vec\lam^0\cdot \vec\lam^a=0$ or $\tilde\lam^a=0$.
Note that the metric $g^{ab}$ in this case is different from eq.\,(\ref{metric-T}) as
\ba\label{metric-gen}
g^{ab}:=|\vec\lam^0|^2(\vec\lam^a\cdot\vec\lam^b)
-(\vec\lam^0\cdot\vec\lam^a)(\vec\lam^0\cdot\vec\lam^b)\,.
\ea
From the discussion above, we can conclude that
the nonabelian $(2,0)$ theory can have an additional total derivative term of the form (\ref{td}) in its Lagrangian,
and that the $F\wedge F\wedge F$ term in D$p$-brane Lagrangian
can be derived from this term.
Here we should remember again that
Lagrangian of the nonabelian $(2,0)$ theory is not defined properly at this stage,
but this discussion is still meaningful,
since the problematic self-dual 2-form field $B_{\mu\nu}$ doesn't appear here at all.
Further justification of this result from the viewpoint of T-transformation will be done in \S\,\ref{sec:U-duality}.

\subsection{Kaluza-Klein monopoles}

For completeness of our discussion, 
we now comment on type IIA/IIB Kaluza-Klein monopoles reproduced from the nonabelian $(2,0)$ theory.

\sss{Type IIA KK monopoles}

It is known that type IIA KK monopoles can be obtained from type IIB NS5-branes
by taking T-duality for a transverse direction~\cite{Eyras:1998hn}.
Therefore, in this case, we use a generalized loop algebra 
$\{T^i_\vm,u_{0,1,2},v^{0,1,2}\}$ defined by eq.\,(\ref{gla}).
Then we put VEV's into $u_{0,1,2}$-component fields as
\ba
X^{I0}=\lam^0 \delta^I_{10}\,,\quad
C^{\mu 1}=\lam^1 \delta^\mu_5\,,\quad
X^{I2}=\lam^2 \delta^I_9\,,\quad
\text{otherwise}=0\,.
\ea
This setup can be generalized into the case where these VEV's are not perpendicular
to each other, but all the following results remain the same.
Finally, we redefine the fields in a similar way to eq.\,(\ref{redef-B}) as
\ba\label{redef-KKA}
\hat \Phi_i(x,y_1,y_2)=\sum_\vm \Phi_{(i\vm)}(x)e^{-i\vm\cdot\vy}\,,\quad
\hat A_{\mu,i}(x,y_1,y_2)=\sum_\vm A^1_{\mu(i\vm)}(x)e^{-i\vm\cdot\vy}\,,\quad
\cdots\,.
\ea
As a result, we obtain the EOM's of the same form as type IIB NS5-brane case in \S\,\ref{sec:NS5B},
except that of the scalar field $\hat X^9$
\ba
\hat D_\alp^2 \hat X^9 + \hat D_{y_1}^2 \hat X^9 
- (\lam^0)^2\lam^1\lam^2\hat D_{y_1}\partial_{y_2} \hat C^5
\!&=&\! 0\,,
\ea
which has an additional term with a $y_2$ derivative, compared with eq.\,(\ref{X-B}). 
We should remember that a factor like $\partial_{y_2} \hat C^\mu$ never appear 
in the previous discussions.
From the viewpoint of Lorentz invariance for the condition $\partial_\mu C^\nu_{(i\vm)}=0$, it is natural here to impose
$\partial_{y_2} \hat C^5=0$, or equivalently, $C^5_{(i\vm)}\bigr|_{m_2\neq 0}=0$.
This, of course, does not break gauge symmetry nor supersymmetry.
After imposing this, the final result does not contain any $y_2$ derivatives,
so this $y_2$ direction becomes isometry.
In fact, it must correspond to Taub-NUT isometry direction.
Therefore, we can integrate out the $y_2$ dependence from 
all the redefined fields (\ref{redef-KKA}),
and then we obtain the 6-dim worldvolume fields in type IIA KK monopole theory
which depend on only $x^{0,\cdots,4}$ and $y_1$ coordinates.

The field contents of this theory are
three embedding scalars $\hat X^{6,7,8}$, 
a 1-form field $\hat A_\umu$,
a 0-form field $\hat X^9$ and a fermion $\hat\Psi$.
Therefore, they are exactly reproduced from the nonabelian $(2,0)$ theory
only by specializing the scalar field $\hat X^9$.

\sss{Type IIB KK monopoles}

On the other hand, type IIB KK monopoles can be obtained from type IIA NS5-branes
by taking T-duality for a transverse direction~\cite{Eyras:1998hn}.
Therefore, in this case, we use a generalized loop algebra
$\{T^i_m,u_{0,1},v^{0,1}\}$ defined by eq.\,(\ref{gla}) or (\ref{loop-B}).
Then we put VEV's into $u_{0,1}$-component fields as
\ba\label{VEV-KKB}
X^{I0}=\lam^0 \delta^I_{10}\,,\quad
X^{I1}=\lam^1 \delta^I_9\,,\quad
\text{otherwise}=0\,.
\ea
Similarly, even if we make these VEV's not perpendicular, 
the following results are unchanged.
Finally, we redefine the field in a similar way to eq.\,(\ref{redef-B}) as
\ba\label{redef-KKB}
\hat \Phi_i(x,y)=\sum_m \Phi_{(im)}(x)e^{-imy}\,,\quad
\cdots\,.
\ea
As a result, at this time, we obtain the EOM's of the same form as
type IIA NS5-brane case (\ref{EOM-NS5A}), except that of the scalar field $\hat X^9$
\ba
\partial_\mu^2 \hat X^9 - (\lam^0)^2[\hat C^\mu,[\hat C_\mu, \hat X^9]
+i(\lam^0)^2\lam^1[\hat C_\mu, \partial_y \hat C^\mu]
\!&=&\! 0\,,
\ea
which has an additional term with a $y$ derivative.
By similar discussion to type IIA KK monopole case,
it is natural to impose $\partial_{y} \hat C^\mu=0$ to eliminate the $y$ derivative,
and to regard the $y$ direction as Taub-NUT isometry direction.
Therefore, we can integrate out the $y$ dependence from all the redefined fields
(\ref{redef-KKB}),
and then we obtain 6-dim worldvolume fields in type IIB KK monopole theory
which depend on only $x^{0,\cdots,5}$ coordinates.
The field contents of this theory are
three embedding scalars $\hat X^{6,7,8}$,
a self-dual 2-form field $\hat B_{\mu\nu}$,
two 0-form fields $\hat X^{9,10}$
and a fermion $\hat\Psi$.
Therefore, they are exactly reproduced from the nonabelian $(2,0)$ theory
only by specializing the scalar fields $\hat X^{9,10}$.

It is also known that type IIB KK monopole theory must be invariant 
under S-duality transformation.
In our setup, 
this transformation corresponds to the interchange of 
VEV's $X^{I0}$ and $X^{I1}$, as we will see in \S\,\ref{sec:U-duality}.
Since $C^\mu$-field has no dynamical degrees of freedom,  
we can regard the resultant theory as practically the simple copies of free theory,
just as we discussed in \S\,\ref{sec:NS5A}.
Therefore, all the interaction terms are negligible,
and then we can see that S-self-duality of type IIB KK monopole is trivially satisfied. 
If one wants to reproduce S-self-duality including the interaction terms,
some generalization of the nonabelian $(2,0)$ theory must be needed.

\subsection{Role of $C^\mu$-field}
\label{sec:C}

Let us make short comments on $C^\mu$-field here.
This field is a nondynamical auxiliary field, since it never has the kinetic term.
Moreover, it seems conveniently introduced instead of a dimensionful parameter
in order to make interaction terms appear in the theory,
since any dimensionful parameters cannot exist in M5-brane system in flat background.

However, let us now try to find some physical meanings of this field.
In fact, it seems related to
the gauge fixing condition for the general coordinate transformation symmetry on the M5-brane worldvolume as
\ba\label{fix}
X^\mu(\sig)=\sig^\mu{\bf 1}+C^\mu_a(\sig)\, T^a\,,
\ea
under the condition $D_\mu C^\nu_a=0$.
Here $\sig^\mu$ are worldvolume coordinates and
${\bf 1}$ is a trivial element,
satisfying $[{\bf 1},T^a,T^b]=0$ and
$\langle {\bf 1}, {\bf 1}\rangle=1$.
It corresponds to the center-of-mass mode in brane system
which is decoupled from the theory.
In the case of generalized loop algebra (\ref{gla}), for example,
$\bf 1$ is equivalent to $T^0_{\vec 0}$.

This discussion suggests that we can regard $[C^\mu,\star,\star\,]$
as $[X^\mu,\star,\star\,]$. 
This identification must be natural:
As we saw in \S\,\ref{sec:Dp} and \S\,\ref{sec:NS5},
putting a VEV for $u$-component of $C^\mu$-field 
means the compactification for one of $x^\mu$-directions,
while putting a VEV for $u$-component of $X^I$-field 
means the compactification for one of $x^I$-directions.
Therefore, it seems very natural to expect that $C^\mu$-field is related to $X^\mu$.



Moreover, we consider in \S\,\ref{sec:genV} that
gauge field $A_{\mu,ab}$ and $C^\mu_a$-field play 
the complementary roles of $X^\mu$.
In fact, in eq.\,(\ref{X5}), we have treated 
the gauge field $A^0_{\mu(i\vm)}$ as $X^\mu_{(i\vm)}$,
while the field $C^{\mu A}$ as $X^\mu_{u_A}$.
The former is natural from the viewpoint of dimensional reduction
where a higher dimensional gauge field is decomposed into a lower dimensional
gauge field and transverse scalars.
However, the latter seems unusual and very interesting.
This makes us again suppose that $C^\mu$-field is related to $X^\mu$.

If the identification (\ref{fix}) is correct,
the condition $D_\mu C^\nu_a=0$ can be regarded as
a gauge fixing for a part of general coordinate transformation symmetry,
which assures that the factor $D_\mu X^\nu_a$ doesn't appear in Lagrangian.
Therefore, in order to check our assumption, we need to write down 
DBI-like action for generalization of the nonabelian $(2,0)$ theory,
since such factors should appear in it. We hope to discuss it in the future.

\section{Discussion on U-duality}
\label{sec:U-duality}

In \S\,\ref{sec:Dp} and \S\,\ref{sec:NS5},
we show that the 
D$p$-brane and NS5-brane theories can be obtained
from the nonabelian $(2,0)$ theory.
Strictly speaking, however, they are only (part of) super Yang-Mills theories,
which are low energy effective theories of the brane systems.
Then in this section, as a further justification of our discussion,
we study whether our results reproduce the expected U-duality relation among M5-branes,
D$p$-branes and NS5-branes.
This must be a highly nontrivial check 
for the nonabelian $(2,0)$ theory as a formulation of M5-brane system.

\subsection{D5-branes on $S^1$}
\label{sec:U-D5}

We start with the simplest case.
This corresponds to the $d=1$ case in \S\,\ref{sec:Dp}.
The notation for VEV's $\vec\lam^A$ is defined in eq.\,(\ref{lam0}) and (\ref{lama}).

\paragraph{T-duality}

For simplicity, only in this and next paragraphs,
let us assume $\vlam^0\perp \vlam^1$.
As we mentioned,
putting the VEV $\vlam^0$ means the compactification of M-theory direction
with the radius
\ba
R_0=|\vlam^0|\,.
\ea
Similarly, 
putting a VEV $\vlam^1$ must imply the compactification
of another direction with the radius $R_1=|\vlam^1|$ before taking T-duality.
Then we have D4-brane worldvolume theory with string coupling~\cite{Lambert:2010wm}
\ba
g_s=g_{YM}^2l_s^{-1}=|\vlam^0|l_s^{-1}
\ea
where $l_s$ is the string length, satisfying $l_p^3=g_sl_s^3$.
In \S\ref{sec:Dp}, D5-brane theory is obtained,
since we take T-duality for the $\vlam^1$ direction (by field redefinition).
After taking T-duality, the compactification radius is
\ba
\tilde R_1=\frac{l_s^2}{R_1}=\frac{l_p^3}{|\vec\lam^0||\vec\lam^1|}\,,
\ea
which is consistent with the metric component $g^{11}$ on the torus $S^1$ (\ref{metric-T}).
From the kinetic term for gauge field in Lagrangian (\ref{Lag}),
the string coupling in this theory can be read as
\ba\label{gs}
g_s'=g_{YM}^{\prime\,2}l_s^{-2}
=\frac{|\vlam^0|^2}{|\vlam^0||\vlam^1|}\frac{l_p^3}{R_0\,l_s^2}
=\frac{|\vlam^0|}{|\vlam^1|}\,,
\ea
which is compatible with the expected result from string duality, namely
$g_s'=g_s l_s/ \tilde R_1=R_0/R_1$.
Therefore, we can conclude that T-duality relation is exactly reproduced.

\paragraph{S-duality}
We continuously assume $\vlam^0\perp \vlam^1$ in this paragraph.
In \S\ref{sec:NS5B}, we discuss the worldvolume theory on type IIB NS5-branes.
From the kinetic term for gauge field in Lagrangian (\ref{Lag-B}),
we can read off the string coupling in this theory as
\ba\label{gs2}
g_s''=g_{YM}^{\prime\prime\,2}l_s^{-2}
=\frac{|\vlam^1|^2}{|\vlam^0||\vlam^1|}\frac{l_p^3}{R_0\,l_s^2}
=\frac{|\vlam^1|}{|\vlam^0|}\,.
\ea
This is exactly the inverse of string coupling in D5-brane theory (\ref{gs}),
so we can conclude that S-duality relation is successfully reproduced.
Moreover,
we can find that S-duality is realized as a part of $SL(2,{\mathbb Z})$ transformation
of VEV's
\ba
\vlam^0 \to -\vlam^1\,,\quad
\vlam^1 \to \vlam^0\,.
\ea

\paragraph{T-transformation}

We consider this transformation in \S\,\ref{sec:genV}.
By comparing the setting of VEV's after transformation (\ref{VEV2}) with
the original one (\ref{VEV}), 
we can find that this transformation is identified with
another part of $SL(2,{\mathbb Z})$ transformation of VEV's
\ba\label{T-trans}
\vlam^0\to \vlam^0\,,\quad
\vlam^1\to \vlam^1+n\vlam^0\,.
\ea
Interestingly enough, it is related to
automorphism of Lie 3-algebra~\cite{Kobo:2009gz}
\ba
&& u_0 \rightarrow u_0 - n u_1 \,,\quad
u_1 \rightarrow u_1 \,,\nn\\
&& v^0 \rightarrow v^0 \,,\quad
v^1 \rightarrow v^1+n v^0 \,,
\ea
that is,
this transformation changes neither structure constant nor metric
of Lie 3-algebra.
The relation between them can be understood as the redefinition of ghost fields
\ba
X^M
= X^{M0} u_0 +X^{M1} u_1+\cdots
= X^{M0} (u_0-n u_1) + (X^{M1}+ n X^{M0})\, u_1+\cdots\,,
\ea
where $M=(\mu,I)$ and $X^{\mu A}:=C^{\mu A}$ as in eq.\,(\ref{X5}).
Of course, there is no reason that the parameter $n$ must be quantized
at the classical level,
but it is still interesting that part of the duality transformation
comes from the automorphism of Lie 3-algebra.

It is well known that this transformation (\ref{T-trans}) causes
the change of axion field $C_{(0)}$, which appears in D5-brane Lagrangian
as a Chern-Simons term $C_{(0)}\wedge F_{(2)} \wedge F_{(2)}\wedge F_{(2)}$.
Therefore, the value of $C_{(0)}$ field can be read from eq.\,(\ref{Lag-CS}) as
\ba
C_{(0)}
=\frac{|\vlam^0|(\vlam^0\cdot\vlam^1)}{3!\,2\pi l_p^3}
=\frac{\tau_1}{3!\,2\pi}\frac{|e|^3}{l_p^3}\,,
\ea
and the inverse of string coupling can be read from eq.\,(\ref{Lag}) as
\ba
g_s^{-1}=\frac{|\vlam^0|\sqrt{g^{11}}}{2\pi l_p^3}
=\frac{\tau_2}{2\pi}\frac{|e|^3}{l_p^3}\,,
\ea
where we define the new basis $\{\ve^{\,0},\ve^{\,1}\}$ as
\ba
\vlam^0=\ve^{\,0}\,,\quad
\vlam^1=\tau_1\ve^{\,0}+\tau_2\ve^{\,1}\,;\quad
\ve^{\,0}\cdot \ve^{\,1}=0\,,\quad
|\ve^{\,0}|=|\ve^{\,1}|=:|e|\,.
\ea
In this basis, T-transformation is written as
$\tau_1\to\tau_1+n$,
$\tau_2\to\tau_2$.
Therefore, this result shows that T-transformation is also perfectly reproduced in our discussion.

\paragraph{Taylor's T-duality}

This transformation~\cite{Taylor:1996ik} interchanges D5- and D4-branes,
and corresponds to the different identification of $T^i_m$-component fields
in our discussion.
To obtain D5-brane system, we constructed 6-dim field $\hat X^I(x,y)$
from the component fields $X^I_{(im)}(x)$ by Fourier transformation (\ref{redef}).
On the other hand, one can interpret $X^I_{(im)}(x)$
as the 5-dim fields and the index $m\in {\mathbb Z}$
as open string modes which interpolate mirror images of
a point in $T^1={\mathbb R}/{\mathbb Z}$.
In this way, Taylor's T-duality transformation ${\mathbb Z}_2$
is reproduced.


\paragraph{Summary}

As we already mentioned, S-duality and T-transformation can be written as
the $SL(2,{\mathbb Z})$ transformation of VEV's
\ba
\left(
\begin{array}{c}
\vec \lambda^1\\ \vec \lambda^0
\end{array}
\right)
\rightarrow \left(
\begin{array}{cc}
a & b\\ c& d
\end{array}
\right)
\left(
\begin{array}{c}
\vec \lambda^1\\ \vec \lambda^0
\end{array}
\right)\,,
\ea
which is equivalent to the transformation of the moduli parameter $\tau:=\tau_1+i\tau_2$
\ba
\tau \rightarrow \frac{a\tau +b}{c\tau +d}\,.
\ea
In fact, S-duality $\tau \rightarrow -1/\tau$ is given as $(a,b,c,d)=(0,1,-1,0)$,
while T-transformation $\tau\rightarrow \tau+n$ is given as $(a,b,c,d)=(1,n,0,1)$.
It is well known that any element of $SL(2,\mathbb{Z})$ transformation can be composed
as combination of these two kinds of transformation.

As a result, together with Taylor's T-duality,
it is finally shown that
the whole of U-duality transformation in the case of D5-branes on $S^1$
(or M-theory on $T^2$)
\ba
SL(2,\mathbb{Z}) \bowtie \mathbb{Z}_2
\ea
is completely reproduced in our discussion,
where the first factor is described by the rotation of VEV's and
the second factor is described by the different representation of the field theory.
Here, the symbol $\bowtie$ denotes the product group defined by the two noncommuting subgroups.

\subsection{D$p$-branes on $T^{p-4}$ $(p\geq 5)$}

Finally, we discuss the U-duality in general $d\geq 1$ cases in \S\,\ref{sec:Dp}.
In these cases, we consider M-theory compactified on $T^{d+1}$ (where $d=p-4$).
This theory has U-duality group 
\ba\label{Ed+1}
E_{d+1}(\mathbb{Z})=SL(d+1,\mathbb{Z})\bowtie SO(d,d;\mathbb{Z})
\ea
and its moduli parameters 
take values in $E_{d+1}/H_{d+1}$,
where $H_{d+1}$ is the maximal compact subgroup of $E_{d+1}$.
(See {\em e.g.}~\cite{Obers:1998fb} for a review.)

Now let us read off the values of these moduli in D$p$-brane case from our results.
For readability, we set $l_p=1$ again in the following.
First, the metric on the torus $T^d$ (\ref{metric-gen}) is
\ba\label{grav}
g^{ab}=|\vec\lam^0|^2(\vec\lam^a\cdot\vec\lam^b)
-(\vec\lam^0\cdot\vec\lam^a)(\vec\lam^0\cdot\vec\lam^b)\,,
\ea
where $a,b=1,\cdots,d$.
Secondly, the Yang-Mills coupling (\ref{Lag}) is
\ba
g_{YM}^2=\frac{(2\pi)^d|\vec\lam^0|}{\sqrt{g}}\,,
\ea
where $g:=\det g^{ab}$.
%
Finally, we read off the value of R-R $(d-1)$-form field $C_{(d-1)}$.
This field may appear in D$p$-brane Lagrangian as a Chern-Simons term
$C_{(d-1)}\wedge F_{(2)}\wedge F_{(2)}\wedge F_{(2)}$.
Therefore, this can be read from eq.\,(\ref{Lag-CS}) as
\ba\label{Cp-5}
C_{(d-1)}=
\frac{|\lam^0|(\vec\lam^0\cdot \vec\lam^a)}{6(2\pi)^d(d-1)!}
\frac{\sqrt g}{\sqrt{g^{aa}}}\,,
\ea
where no sum is taken on the index $a$.
This represents the components of $C_{(d-1)}$ with the indices
$1\,2 \cdots \hat a \cdots d$, {\em i.e.}~except $a$.

Therefore, the number of moduli written by VEV's (\ref{grav})--(\ref{Cp-5}) is
\ba
\frac12 d(d+1)+1+d=\frac12 (d+1)(d+2)\,.
\ea
This coincides with the number of parameters in $G^{AB}:=\vec\lam^A\cdot\vec\lam^B$,
which is transformed under $SL(2,\mathbb{Z})$ transformation
\ba
\vec\lam^A\to\vec\lam'^A:=\Lambda^A{}_B\vec\lam^B\,;\quad
\Lambda^A{}_B\in SL(d+1,\mathbb{Z})\,.
\ea
This means that
our discussion correctly reproduces the $SL(d+1,\mathbb{Z})$ symmetry
as the first factor of U-duality (\ref{Ed+1}),
and that $G^{AB}=G^{AB}(g^{ab},\,g_{YM}^2,\,C_{(d-1)})$ gives the moduli parameter
which is transformed covariantly under the $SL(d+1,\mathbb{Z})$ transformation.

The second factor $SO(d,d;\mathbb{Z})$ of U-duality (\ref{Ed+1}) can be also
reproduced.
It consists of the permutation of T-duality directions,
Taylor's T-duality transformation,
and the shift of the value of NS-NS 2-form field.
The first one can be seen trivially in our setup,
and the second one is reproduced in a similar way to the $d=1$ case.
The third one is rather nontrivial.
The NS-NS 2-form field $B_{ab}$ can be
introduced as the deformation of Lie 3-algebra~\cite{Ho:2009nk}
\ba\label{NSNS2}
{}[u_0,u_a,u_b]=B_{ab}T_{\vec 0}^0
\ea
instead of ordinary generalized loop algebra (\ref{gla}),
since it provides the noncommutativity on the torus $T^d$.
It is interesting that some part of moduli (\ref{grav})--(\ref{Cp-5})
are described in terms of VEV's,
while another part comes from the structure constant of Lie 3-algebra.

However, this is not the end of the story.
The U-duality group is a product of these {\em noncommuting} subgroups,
and so unfortunately,
the whole moduli space of U-duality cannot be described by only
the moduli parameters obtained above.
In the following, we check the dimension of moduli space, and
discuss what kinds of parameters are lacked in our setup.
In fact, in the $d\geq 3$ cases, some missing parameters exist.

\paragraph{D5-branes $(d=1)$}

M-theory compactified on $T^2$ is considered.
The moduli space in this case is
$\bigl(SL(2)/U(1)\bigr)\times \mathbb{R}$
which gives 3 parameters.
They correspond to $g^{11}$, $\phi$ and $C_{(0)}$.

\paragraph{D6-branes $(d=2)$}

M-theory compactified on $T^3$ is considered.
The moduli space in this case is
$\bigl(SL(3)/SO(3)\bigr)\times \bigl(SL(2)/U(1)\bigr)$
which gives 7 parameters.
They correspond to $g^{ab}$, $B_{ab}$, $\phi$ and $C_{(1)}$
which transform in the ${\bf 3} + {\bf 1} + {\bf 1} + {\bf 2}$
representations of $SL(2)$.


\paragraph{D7-branes $(d=3)$}

M-theory compactified on $T^4$ is considered.
The moduli space in this case is $SL(5)/SO(5)$
which gives 14 parameters.
They correspond to
$g^{ab}$, $B_{ab}$, $\phi$, $C_{(2)}$ and $C_{(0)}$
which transform in the ${\bf 6} + {\bf 3} + {\bf 1} + {\bf 3} + {\bf 1}$
representations of $SL(3)$.

R-R 0-form field $C_{(0)}$ is lacked in our discussion.
This field causes the Chern-Simons interaction
$C_{(0)}\wedge F_{(2)}\wedge F_{(2)}\wedge F_{(2)}\wedge F_{(2)}$
which cannot be derived 
in a similar way to \S\,\ref{sec:genV}.
Therefore, in order to include this parameter, we might need to
consider the 
nontrivial backgrounds.
For the missing parameters below, similar discussions would be made.

\paragraph{D8-branes $(d=4)$}

M-theory compactified on $T^5$ is considered.
The moduli space in this case is
$SO(5,5)/\bigl(SO(5)\times SO(5)\bigr)$
which gives 25 parameters.
They correspond to
$g^{ab}$, $B_{ab}$, $\phi$, $C_{(3)}$ and $C_{(1)}$
which transform in the
${\bf 10} + {\bf 6} + {\bf 1} + {\bf 4} + {\bf 4}$
representation of $SL(4)$.
R-R 1-form field $C_{(1)}$ is lacked in our discussion.

\paragraph{D9-branes $(d=5)$}

M-theory compactified on $T^6$ is considered.
The moduli space in this case is $E_6/USp(8)$
which gives 42 parameters.
They correspond to
$g^{ab}$, $B_{ab}$, $\phi$, $C_{(4)}$, $C_{(2)}$ and $C_{(0)}$
which transform in the
${\bf 15} + {\bf 10} + {\bf 1} + {\bf 5} + {\bf 10} + {\bf 1}$
representations of $SL(5)$.
R-R 2-form and 0-form field $C_{(2)}$, $C_{(0)}$ are lacked in our discussion.

\section{Conclusion and discussion}
\label{sec:Concl}

In this paper, we derive the super Yang-Mills action for
D$p$-brane and type IIB NS5-brane systems on a torus $T^d$ ($d=p-4$)
from the nonabelian $(2,0)$ theory.
We use the generalized loop algebra as an example of Lie 3-algebra,
which has general $d+1$ negative norm generators.
The ghost fields from these generators can be removed
by putting VEV's for them through the novel Higgs mechanism.
In the resultant theories,
the Yang-Mills coupling $g_{YM}^2$ and all the moduli of the compactified torus $T^d$
can be exactly written in terms of these VEV's.
From the discussion on total derivative terms,
we see that R-R $(d-1)$-form field $C_{(d-1)}$ is also given by the VEV's.
For D5-brane ($d=1$) case, the moduli parameters thus obtained are enough to
realize the whole of U-duality relation through $SL(2,\mathbb{Z})$ transformation
of the VEV's.
Especially, S-duality between D5-branes and type IIB NS5-branes is clearly reproduced.
For higher $d$ cases, 
these parameters only compose a subgroup of U-duality
transformation $SL(d+1,\mathbb{Z})$, and so
extra R-R fields are needed in order to recover all of
the moduli parameters.
This might be achieved by generalizing the nonabelian $(2,0)$ theory
in nontrivial backgrounds.

For the future direction, therefore, the generalization in background fields is one of the important subjects.
For a single M5-brane in $C_{(3)}$ background, the action has already derived
in the context of BLG theory~\cite{Ho:2008ve}.
In this M5-brane theory, the term $[X^I,X^J,X^K]^2$ appears in the Lagrangian.
If the nonabelian $(2,0)$ theory has such a term, the dimensionful parameters
must be introduced.
Then we expect that this term can be naturally added to the theory
by considering the $C_{(3)}$ background as the parameters.
Also, we should remember that the resultant type IIB NS5-brane Lagrangian doesn't
have interaction terms.
In order to derive all the terms in super Yang-Mills action,
we need again to add some terms into the theory,
which might come from the generalization of backgrounds.
Moreover, we expect that such a generalization enables us to reproduce S-self-duality 
including the interaction terms in type IIB Kaluza-Klein monopole theory.

Clarifying the meaning of the $C^\mu$-field is also an interesting topic.
Some assumptions at this moment were mentioned in \S\,\ref{sec:C},
but the further discussion should be done.
Especially, when we discuss type IIA/IIB NS5-brane and Kaluza-Klein monopole systems,
this field remains in the final form.
Therefore, we expect that such discussion will provide us  
a clearer understanding of these systems.

The search for more appropriate multiple M5-brane theory is also important.
It is problematic that
in some sense
the nonabelian $(2,0)$ theory is automatically reduced to D4-brane theory,
and that it seems not to represent the full multiple M5-brane dynamics.
This situation may be similar to Lorentzian BLG theory~\cite{Mukhi:2008ux,Ho:2008ei}. 
As shown in~\cite{Honma:2008jd}, this theory is emerged from the scaling limit of ABJM theory~\cite{Aharony:2008ug},
and thus it does not include the full M2-brane dynamics.
Therefore, in analogy with this relation,
it is just conceivable that 
a certain scaling limit of some 6-dim theory 
which correctly describes the multiple M5-branes 
may provide this nonabelian $(2,0)$ theory. 

Finally, writing down the Lagrangian remains a serious problem.
It is obvious that some modification of the theory must be needed.
Until this problem is solved,
we cannot make clear the relation to the BLG Lagrangian~\cite{Bagger:2007jr}
and the covariant action for single M5-brane (PST action)~\cite{Bandos:1997ui}.
Therefore, exploring this problem is indispensable for the covariant formulation of the multiple M5-branes.

\subsection*{Acknowledgment}

We would like to thank Pei-Ming Ho, Yutaka Matsuo and Tsunehide Kuroki 
for valuable discussions and comments.
S. S. would like to thank the hospitality of particle physics group
at Institute of Advanced Study.
Y. H. would like to thank National Taiwan University and National Tsing Hua University for their hospitality.
S. S. is partially supported by
Grant-in-Aid for Scientific Research (B) \#19340066.
Y. H. is partially supported by
Grant-in-Aid for JSPS Fellows \#22$\cdot$3677.

\appendix

\end{document}